\documentclass{ws-ijmpa}
\usepackage[super,compress]{cite}
\usepackage{amsmath}
\usepackage{hyperref}
\hypersetup{colorlinks,urlcolor=black,citecolor=black,linkcolor=black,filecolor=black}
\usepackage{breakurl}
\begin{document}
	\markboth{Authors' Names}{Instructions for typing manuscripts (paper's title)}
	
	%
	\catchline{}{}{}{}{}
	%
	
	\title{Strong Decays and Spin-Parity Assignments of Low-Lying Singly Charmed Baryons
	}
	
	\author{Pooja Jakhad$^{*}$, Juhi Oudichhya$^{\dagger}$ and Ajay Kumar Rai$^{\ddagger}$
	}
	
	\address{Depertment of physics\\Sardar Vallabhbhai National Institute of Technology \\
		Surat 395007, Gujarat, India 
		\\
		*poojajakhad6@gmail.com
		\\
		$\dagger$juhioudichhya01234@gmail.com
		\\
		$\ddagger$raiajayk@gmail.com}

	\maketitle
	
	\begin{history}
		\received{Day Month Year}
		\revised{Day Month Year}
	\end{history}
	
\begin{abstract}
	
In this study, we analyze the strong decays of singly charmed baryons into another singly charmed baryon and a light pseudo scalar meson by means of  heavy hadron chiral perturbation theory (HHChPT) which combines heavy quark symmetry with chiral symmetry.  HHChPT worked excellently for describing the strong decays of $1S$-wave charmed baryons. The strong decay widths of well-established singly charmed baryons are well reproduced. Further, the strong decay width of $1P$ states are calculated in this approach by using the calculated masses in the relativistic flux tube (RFT) model. By analysing the mass spectra from the RFT model with strong decays in HHChPT, we investigate the spin-parity assignments. We identify $\Sigma_{c}(2800)$ and $\Xi_{c}(2939)$ as $1P$-wave states having $J^{P}=\frac{3}{2}^{-}$.

\end{abstract}
	
	\ccode{PACS numbers:14.20.Lq}
	

\section{Introduction}

Single charmed baryons, which consist of one charm quark and two lighter quarks (up, down, or strange), are crucial for studying quantum chromodynamics (QCD) and the strong interaction. These baryons provide valuable insights into the behaviour of heavy quarks within baryons, offering essential data for testing and refining theoretical models. In recent years, the experimental investigation of low-lying single charmed baryons has advanced significantly, driven by improvements in particle detection and accelerator technologies at the LHC, Belle, BaBar, etc..	

For the $\Lambda_c$ and $\Xi_c$ baryon families, states corresponding to the $1S$- and $1P$-wave have been well identified. Conversely, for the $\Sigma_c$, $\Xi_c'$, and $\Omega_c$ baryon families, only the states associated with the $1S$-wave have been established. Experimental studies have identified a several  low-lying states of the $\Sigma_c$, $\Xi_c'$, and $\Omega_c$ baryons; however, their spin-parity assignments are still unknown. 
In Ref.\cite{Oudichhya:2023awb}, the author assigns the spin-parity of these  experimental states utilizing Regge phenomenology.  From both experimental and theoretical perspectives, it is crucial to determine the quantum numbers of these new states and to comprehend their properties.
{Numerous studies have investigated the strong decay widths of charmed baryons, such as the $^{3}P_{0}$ model \cite{ PhysRevD.107.034031, RamirezMorales:2022lsy, Bijker:2020tns, PhysRevD.96.114009}, 
	the chiral quark model\cite{PhysRevD.96.116016, PhysRevD.95.116010, PhysRevD.95.014023}, the non relativistic quark model\cite{PhysRevD.72.094022, Yao:2018jmc}, the QCD sum rule\cite{PhysRevD.95.094008, Agaev:2017lip, Zhu:2000py}, the MIT bag model \cite{Hwang:2006df}, and the heavy hadron chiral perturbation theory\cite{Gandhi:2019xfw, Gandhi:2019bju}. As various theoretical approaches yield different predictions about the spin parity for experimentally observed states, it is important to do more theoretical investigations and compare them with experimental data in order to identify them.} 

	
In our previous study, we calculated the mass spectra of all singly heavy baryons using the RFT model and assigned spin-parity quantum numbers to many experimentally observed states \cite{PhysRevD.108.014011,Jakhad:2024wpx,jakhad2023pos, Jakhad:2024whd, Jakhad:2024fin}. In this study, we aim to validate these spin-parity assignments for low-lying states such as $1S$ and $1P$ states through strong decay analysis of singly charmed baryons, as the reliability of spin-parity assignments is enhanced when both mass spectra and decay analysis are considered. 
	
The paper is structured as follows: In Section II, we outline the method for calculating the strong decay widths. Further, by analysing the mass spectrum and strong decay widths, we discuss the $J^{P}$ assignments for the experimentally observed states. In Section III, we present our conclusions.

\section{Strong decay}

\begin{table}[h]
	\tbl{{Masses of $1S$ and $1P$ states of singly charmed baryons.}}
	{\begin{tabular}{cccc}
			\toprule
			States	&\textit{$|nL,J^P\rangle_{j}$}& $M_{th}$\cite{PhysRevD.108.014011} &  $M_{exp}$\cite{Workman:2022ynf} \\
			&&(MeV) & (MeV) \\   
			\hline
			$\Lambda_{c}(\frac{1}{2}^{+})$&\textit{$|1S,\frac{1}{2}^+\rangle_{j=0}$}&2286.5 &2286.46$\pm$0.14 \\
			$\Lambda_{c1}(\frac{1}{2}^{-})$&\textit{$|1P,\frac{1}{2}^-\rangle_{j=1}$}&2592.3 &2592.25$\pm$0.28 \\
			$\Lambda_{c1}(\frac{3}{2}^{-})$&\textit{$|1P,\frac{3}{2}^-\rangle_{j=1}$}&2628.1 &2628.10$\pm$0.19 \\
			\hline
			$\Xi_{c}(\frac{1}{2}^{+})$&\textit{$|1S,\frac{1}{2}^+\rangle_{j=0}$}&2470.4 &2470.44$\pm$0.28 \\
			$\Xi_{c1}(\frac{1}{2}^{-})$&\textit{$|1P,\frac{1}{2}^-\rangle_{j=1}$}&2785.7 &2793.90$\pm$0.50 \\
			$\Xi_{c1}(\frac{3}{2}^{-})$&\textit{$|1P,\frac{3}{2}^-\rangle_{j=1}$}&2823.9 &2819.79$\pm$0.30 \\
			\hline
			
			$\Sigma_{c}(\frac{1}{2}^{+})$&\textit{$|1S,\frac{1}{2}^+\rangle_{j=1}$}&2454.0 &2453.97$\pm$0.14 \\
			$\Sigma_{c}^{*}(\frac{3}{2}^{+})$&\textit{$|1S,\frac{3}{2}^+\rangle_{j=1}$}&2518.4 &2518.41$^{+0.22}_{-0.18}$ \\
			
			$\Sigma_{c0}(\frac{1}{2}^{-})$&\textit{$|1P,\frac{1}{2}^-\rangle_{j=0}$}&2727.8 & \\	
			$\Sigma_{c1}(\frac{1}{2}^{-})$&\textit{$|1P,\frac{1}{2}^-\rangle_{j=1}$}&2749.3 & \\
			$\Sigma_{c1}(\frac{3}{2}^{-})$&\textit{$|1P,\frac{3}{2}^-\rangle_{j=1}$}&2800.1 &2801.00$^{+4.00}_{-6.00}$ \\
			
			$\Sigma_{c2}(\frac{3}{2}^{-})$&\textit{$|1P,\frac{3}{2}^-\rangle_{j=2}$}&2872.1 & \\
			
			$\Sigma_{c2}(\frac{5}{2}^{-})$&\textit{$|1P,\frac{5}{2}^-\rangle_{j=2}$}&2908.5 & \\
			
			
			
			\hline
			
			$\Xi_{c}^{'}(\frac{1}{2}^{+})$&\textit{$|1S,\frac{1}{2}^+\rangle_{j=1}$}&2578.7 &2578.70$\pm$0.50 \\		 
			$\Xi_{c}^{*}(\frac{3}{2}^{+})$&\textit{$|1S,\frac{3}{2}^+\rangle_{j=1}$}&2646.2 &2646.16$\pm$0.25 \\
			$\Xi_{c0}^{'}(\frac{1}{2}^{-})$&\textit{$|1P,\frac{1}{2}^-\rangle_{j=0}$}&2873.3 & \\             
			$\Xi_{c1}^{'}(\frac{1}{2}^{-})$&\textit{$|1P,\frac{1}{2}^-\rangle_{j=1}$}&2886.4 &2923.04$\pm$0.35 \\                  
			$\Xi_{c1}^{'}(\frac{3}{2}^{-})$&\textit{$|1P,\frac{3}{2}^-\rangle_{j=1}$}&2937.9 &2938.55$\pm$0.30\\


			$\Xi_{c2}^{'}(\frac{3}{2}^{-})$&\textit{$|1P,\frac{3}{2}^-\rangle_{j=2}$}&2992.9 & \\


			$\Xi_{c2}^{'}(\frac{5}{2}^{-})$&\textit{$|1P,\frac{5}{2}^-\rangle_{j=2}$}&3030.5 & \\
			
			\hline
			$\Omega_{c}(\frac{1}{2}^{+})$&\textit{$|1S,\frac{1}{2}^+\rangle_{j=1}$}&2695.2 &2695.20$\pm$1.70 \\
			$\Omega_{c}^{*}(\frac{3}{2}^{+})$&\textit{$|1S,\frac{3}{2}^+\rangle_{j=1}$}&2765.9 &2765.90$\pm$2.00 \\
			$\Omega_{c0}(\frac{1}{2}^{-})$&\textit{$|1P,\frac{1}{2}^-\rangle_{j=0}$}&3003.2 &3000.41$\pm$0.22 \\
			$\Omega_{c1}(\frac{1}{2}^{-})$&\textit{$|1P,\frac{1}{2}^-\rangle_{j=1}$}&3010.7 &3050.19$\pm$0.13 \\
			$\Omega_{c1}(\frac{3}{2}^{-})$&\textit{$|1P,\frac{3}{2}^-\rangle_{j=1}$}&3062.8 &3065.54$\pm$0.26 \\
			$\Omega_{c2}(\frac{3}{2}^{-})$&\textit{$|1P,\frac{3}{2}^-\rangle_{j=2}$}&3106.6 &3090.10$\pm$0.50\\
			$\Omega_{c2}(\frac{5}{2}^{-})$&\textit{$|1P,\frac{5}{2}^-\rangle_{j=2}$}&3145.3 &3119.10$\pm$1.00\\ 			
			\botrule          	
		\end{tabular}  \label{table1}  }
\end{table}

Heavy hadron chiral perturbation theory (HHChPT), which integrates heavy-quark symmetry and chiral symmetry, offers a highly convenient framework for describing the strong decays of singly charmed baryons into another singly charmed baryon and a light pseudoscalar meson. The general chiral Lagrangian for heavy baryons in HHChPT is constructed using superfields as detailed in Ref. \cite{PhysRevD.50.3295, PhysRevD.56.5483}. This framework employs  two coupling constants, $g_{1}$ and $g_2$, for $P$-wave transitions between $S$-wave baryons, six coupling constants, $h_{2}$ to $h_{7}$, for $S$-wave transitions between $S$-wave and $P$-wave baryons, and eight coupling constants, $h_{8}$ to $h_{15}$, for $D$-wave transitions between $S$-wave and $P$-wave baryons \cite{PhysRevD.55.5851, PhysRevD.56.5483, Kakadiya:2021jtv, Kakadiya:2022zvy}. 

The strong decay width expressions derived from the chiral Lagrangian in this approach can be used to calculate the strong decay widths for the $1S$- and $1P$-wave states of singly charmed baryons, utilizing our calculated masses in Ref. \cite{PhysRevD.108.014011}. In Ref. \cite{PhysRevD.108.014011}, the experimentally observed states of single charmed baryons were  well described by the relativistic flux tube model with heavy quark–light diquark picture of singly charmed baryons. {Our calculated masses, $M_{th}$, are listed in Table (\ref{table1}) alongside the experimentally measured mass, $M_{exp}$.} Here, $n$, $L$, $j$, $J$, and $P$ denote the principal quantum number, orbital angular momentum quantum number, total angular momentum number of the diquark, total spin of the baryon, and parity of the baryon, respectively. In quark-diquark picture of singly charmed baryons, there is only one state with $J^{P}=\frac{1}{2}^{+}$  in $1S$-wave for  $\Lambda_{c}$ and $\Xi_{c}$ baryons  while there are  two states with $J^{P}=\frac{1}{2}^{+}$ and $\frac{3}{2}^{+}$ for each $\Sigma_{c}$, $\Xi_{c}^{'}$ and $\Omega_{c}$ baryons. Further in $1P$-wave  for $\Lambda_{c}$ and $\Xi_{c}$ baryons, there are two possible states for each baryon, named $\Lambda_{c1}(\frac{1}{2}^{-},\frac{3}{2}^{-})$ and $\Xi_{c1}(\frac{1}{2}^{-},\frac{3}{2}^{-})$, while for $\Sigma_{c}$, $\Xi_{c}^{'}$ and $\Omega_{c}$ baryons, there are five possible states for each baryon named $\Sigma_{c0}(\frac{1}{2}^{-})$,  $\Sigma_{c1}(\frac{1}{2}^{-},\frac{3}{2}^{-})$, $\Sigma_{c2}(\frac{3}{2}^{-},\frac{5}{2}^{-})$,  $\Xi_{c0}^{'}(\frac{1}{2}^{-})$, $\Xi_{c1}^{'}(\frac{1}{2}^{-},\frac{3}{2}^{-})$,  $\Xi_{c2}^{'}(\frac{3}{2}^{-},\frac{5}{2}^{-})$, $\Omega_{c0}(\frac{1}{2}^{-})$,  $\Omega_{c1}(\frac{1}{2}^{-},\frac{3}{2}^{-})$, $\Omega_{c2}(\frac{3}{2}^{-},\frac{5}{2}^{-})$.  In the following part of this section, we calculate the strong decay width for these states and discuss the obtained results.

	\subsection{$\Lambda_{c}$}
	\begin{table}[h]
		\tbl{{Two-body strong decay widths of $1S$ and $1P$ states of $\Lambda_{c}$ and $\Xi_{c}$ baryons. Decay widths represented by - signify that the strong decay channel is not available for particular states. Partial widths of 0 suggest phase space prohibits the perticular decay channel. The calculated total strong decay widths are compared with the experimental widths listed in the Particle Data Group (PDG) \cite{Workman:2022ynf} for quantum number assignment to the observed states. The results from other theoretical frameworks\cite{PhysRevD.107.034031, PhysRevD.96.116016, PhysRevD.95.094008 } are presented for comparison.}}
		{\begin{tabular}{cccccccc}
				\toprule
				States	& Decay  & Partial & Total  & Exp.\cite{Workman:2022ynf} & Ref.\cite{PhysRevD.107.034031}& Ref.\cite{PhysRevD.96.116016}& Ref.\cite{PhysRevD.95.094008}\\
				&modes&width &width&width& & & \\
				&& (MeV)&(MeV)&(MeV)&(MeV) &(MeV) &(MeV)\\

				\hline
				$\Lambda_{c}^{+}(\frac{1}{2}^{+})$& Weak         &-&-& -& & & \\
				$\Lambda_{c1}^{+}(\frac{1}{2}^{-})$& $\Sigma_{c}\pi$         & 0.92$\pm$0.48 & 0.92 & 2.60$\pm$0.60 & 2$^{+1}_{-1}$& & 32\\
				$\Lambda_{c1}^{+}(\frac{3}{2}^{-})$& $\Sigma_{c}\pi$         &0.095$\pm$0.013 &0.095$\pm$0.013 & $<$0.97 &10$^{+5}_{-5}$ & & 0.96\\
				\hline
				$\Xi_{c}^{+}(\frac{1}{2}^{+})$& Weak         &-&-& -& & & \\
				$\Xi_{c}^{0}(\frac{1}{2}^{+})$& Weak         &-&-& -& & & \\
				$\Xi_{c1}^{+}(\frac{1}{2}^{-})$& $\Xi_{c}^{'}\pi$         & 7.39$\pm$2.86& 7.39& 8.90$\pm$1.00& 3$^{+2}_{-2}$& 6.31& 100\\
				$\Xi_{c1}^{0}(\frac{1}{2}^{-})$& $\Xi_{c}^{'}\pi$         & 7.39$\pm$2.86& 7.39& 10.00$\pm$1.10 & & & \\
				
				$\Xi_{c1}^{+}(\frac{3}{2}^{-})$& $\Xi_{c}^{'}\pi$        &0.56$\pm$0.10 & 4.65$\pm$1.58 & 2.43$\pm0.26$ & 5$^{+2}_{-2}$&2.11 &30 \\
				&$\Xi_{c}^{*}\pi$  &4.09$\pm$1.58 & & & & & \\
				$\Xi_{c1}^{0}(\frac{3}{2}^{-})$& $\Xi_{c}^{'}\pi$          &0.56$\pm$0.10 &4.65$\pm$1.58 & 2.54$\pm0.25$ & & & \\
				&$\Xi_{c}^{*}\pi$ &4.09$\pm$1.58 & & & & & \\
				\botrule          			& & & 
			\end{tabular} \label{table2}}
	\end{table}

	The $\Lambda_c$ baryon family has been extensively studied in various high-energy physics experiments. So, the $1S$- and  $1P$-wave states of $\Lambda_{c}$ baryon are firmly established in experiment. The ground state, $\Lambda_c^+$, decays primarily via the weak interaction. The $\Lambda_{c}(2595)$ and $\Lambda_{c}(2625)$ were first observed in $\Lambda_{c}^{+}\pi^{+}\pi^{-}$ channel at CLEO\cite{PhysRevLett.74.3331} and ARGUS\cite{ALBRECHT1993227}, respectively. They forms a doublet $\Lambda_{c1}(\frac{1}{2}^{-},\frac{3}{2}^{-})$.
	
	The allowed two body strong decay for  $\Lambda_{c1}(\frac{1}{2}^{-})$ is $\Sigma_{c}\pi$  in $S$-wave, while for $\Lambda_{c1}(\frac{3}{2}^{-})$ is $\Sigma_{c}\pi$  in $D$-wave. In HHChPT, the expression for the two-body strong decay width for these states are given by\cite{PhysRevD.75.014006}:

	\begin{equation}
		\label{eq:1}
		\begin{split}
			\Gamma[	\Lambda_{c1}^{+}(\frac{1}{2}^{-})]&=\Gamma[\Lambda_{c1}^{+}(\frac{1}{2}^{-})\rightarrow\Sigma_{c}^{++}\pi^{-}, \Sigma_{c}^{+}\pi^{0},\Sigma_{c}^{0}\pi^{+}]\\
			&=\frac{h_{2}^{2}}{2\pi f_{\pi}^{2}} \left(\frac{M_{\Sigma_{c}^{++}}}{M_{\Lambda_{c1}^{+}(\frac{1}{2}^{-})}}E_{\pi^{-}}^{2}p_{\pi^{-}}+\frac{M_{\Sigma_{c}^{+}}}{M_{\Lambda_{c1}^{+}(\frac{1}{2}^{-})}}E_{\pi^{0}}^{2}p_{\pi^{0}}+\frac{M_{\Sigma_{c}^{0}}}{M_{\Lambda_{c1}^{+}(\frac{1}{2}^{-})}}E_{\pi^{+}}^{2}p_{\pi^{+}}\right),
		\end{split}
	\end{equation}
	
	\begin{equation}
		\label{eq:2}
		\begin{split}
			\Gamma[\Lambda_{c1}^{+}(\frac{3}{2}^{-})]&=\Gamma[\Lambda_{c1}^{+}(\frac{3}{2}^{-})\rightarrow\Sigma_{c}^{++}\pi^{-}, \Sigma_{c}^{+}\pi^{0},\Sigma_{c}^{0}\pi^{+}]\\
			&=\frac{2 h_{8}^{2}}{9\pi f_{\pi}^{2}} \left(\frac{M_{\Sigma_{c}^{++}}}{M_{\Lambda_{c1}^{+}(\frac{1}{2}^{-})}}p_{\pi^{-}}^{5}+\frac{M_{\Sigma_{c}^{+}}}{M_{\Lambda_{c1}^{+}(\frac{1}{2}^{-})}}p_{\pi^{0}}^{5}+\frac{M_{\Sigma_{c}^{0}}}{M_{\Lambda_{c1}^{+}(\frac{1}{2}^{-})}}p_{\pi^{+}}^{5}\right),
		\end{split}
	\end{equation}
where $p_{\pi}$ and $E_{\pi}$ denote the pion's momentum and energy in the center of mass frame, respectively, with  $f_{\pi}$=132. {In the framework of HHChPT, the coupling constants for the $S$-wave transition is $h_{2}=0.437^{+0.114}_{-0.102}$, and that for $D$-wave transition is $h_{8}=(0.86^{+0.08}_{-0.10})$$*10^{-3}$ MeV$^{-1}$  \cite{PhysRevD.75.014006}. }

{Using the theoretical masses $M_{th}$ calculated in the RFT model, in  Eq. (\ref{eq:1}) and (\ref{eq:2}), we calculate the strong decay width. The uncertainties in coupling constants are transmitted to the strong decay widths.  The partial and total decay widths, including errors, are presented in Table \ref*{table2}, alongside the experimentally measured width\cite{Workman:2022ynf} and results from various theoretical frameworks, such as the $^{3}P_{0}$ model\cite{PhysRevD.107.034031}, the chiral quark model\cite{PhysRevD.96.116016},  and light cone QCD sum rules\cite{PhysRevD.95.094008}.} We observe that the calculated widths  for the $\Lambda_{c1}^{+}(\frac{1}{2}^{-})$ and $\Lambda_{c1}^{+}(\frac{3}{2}^{-})$ states are less than the measured widths for $\Lambda_{c}(2595)$ and $\Lambda_{c}(2625)$. This discrepancy arises because the $\Lambda_{c1}^{+}(\frac{1}{2}^{-})$ and $\Lambda_{c1}^{+}(\frac{3}{2}^{-})$ states can also decay via the resonant decay mode $\Lambda_{c}\pi^{+}\pi^{-}$, while we considered only two-body strong decays. Therefore, our calculated widths for these states are less than the experimental total decay widths of $\Lambda_{c}(2595)$ and $\Lambda_{c}(2625)$.

\subsection{$\Xi_{c}$}

$\Xi_{c}$ baryonic family also contain well established states in $1S$- and $1P$-wave. The ground states, $\Xi_{c}^{+}$ and $\Xi_{c}^{0}$ decay primarily via the weak interaction.

The two $1P$-wave states, $\Xi_{c}(2790)$ and $\Xi_{c}(2815)$, were first observed at CLEO in  $\Xi_{c}^{'}\pi$ and $\Xi_{c}^{*}\pi$ channels, respectively. They form a $1P$-wave doublet, named $\Xi_{c1}(\frac{1}{2}^{-},\frac{3}{2}^{-})$. Since the diquark transition $1^{-}\rightarrow 0^{+}+\pi$ is forbidden, $\Xi_{c1}(\frac{1}{2}^{-},\frac{3}{2}^{-})$ cannot decay into $\Xi_{c}\pi$. The possible strong decay mode for  $\Xi_{c1}(\frac{1}{2}^{-})$ is $\Xi_{c}^{'}\pi$  in $S$-wave, while for $\Xi_{c1}(\frac{3}{2}^{-})$, the decay modes are  $\Xi_{c}^{*}\pi$ in $S$-wave and $\Xi_{c}^{'}\pi$ in $D$-wave. In HHChPT, the expressions for it's decay width are \cite{PhysRevD.75.014006}:
	\begin{equation}
		\label{eq:3}
		\begin{split}
			\Gamma[\Xi_{c1}^{+}(\frac{1}{2}^{-})]&=\Gamma[\Xi_{c1}^{+}(\frac{1}{2}^{-})\rightarrow\Xi_{c}^{'+}\pi^{0},\Xi_{c}^{'0}\pi^{+}]\\
			&=\frac{h_{2}^{2}}{2\pi f_{\pi}^{2}} \left(\frac{1}{4}\frac{M_{\Xi_{c}^{'+}}}{M_{\Xi_{c1}^{+}({\frac{1}{2}}^-)}}E_{\pi^{0}}^{2}p_{\pi^{0}}+\frac{1}{2}\frac{M_{\Xi_{c}^{'0}}}{M_{\Xi_{c1}^{+}({\frac{1}{2}}^-)}}E_{\pi^{+}}^{2}p_{\pi^{+}}\right),\\
		\end{split}
	\end{equation}
	
	\begin{equation}
		\label{eq:4}
		\begin{split}
			\Gamma[\Xi_{c1}^{+}(\frac{3}{2}^{-})]&=\Gamma[\Xi_{c1}^{+}(\frac{3}{2}^{-})\rightarrow\Xi_{c}^{'+}\pi^{0},\Xi_{c}^{'0}\pi^{+},\Xi_{c}^{*+}\pi^{0},\Xi_{c}^{*0}\pi^{+}]\\
			&=\frac{2 h_{8}^{2}}{9\pi f_{\pi}^{2}} \left(\frac{1}{4}\frac{M_{\Xi_{c}^{'+}}}{M_{\Xi_{c1}^{+}(\frac{3}{2}^{-})}}p_{\pi^{0}}^{5}+\frac{1}{2}\frac{M_{\Xi_{c}^{'0}}}{M_{\Xi_{c1}^{+}(\frac{3}{2}^{-})}}p_{\pi^{+}}^{5}\right)\\
			&+\frac{h_{2}^{2}}{2\pi f_{\pi}^{2}} \left(\frac{1}{4}\frac{M_{\Xi_{c}^{*+}}}{M_{\Xi_{c1}^{+}(\frac{3}{2}^{-})}}E_{\pi^{0}}^{2}p_{\pi^{0}}+\frac{1}{2}\frac{M_{\Xi_{c}^{*0}}}{M_{\Xi_{c1}^{+}(\frac{3}{2}^{-})}}E_{\pi^{+}}^{2}p_{\pi^{+}}\right).
		\end{split}
	\end{equation}
Similar expressions to Eq. (\ref{eq:3})-(\ref{eq:4}) can be used to calculate the strong decay widths of the isospin counterparts, $\Xi_{c1}^{0}(\frac{1}{2}^{-})$ and $\Xi_{c1}^{0}(\frac{3}{2}^{-})$. The obtained results are listed in Table \ref{table2}. We observe that the calculated width from HHChPT is in good agreement with the experimentally measured width for $1P$-wave states of $\Xi_{c}$ baryon.

\subsection{$\Sigma_{c}$}
\begin{table}
	\tbl{{Same as in Table \ref{table2}, but for $\Sigma_{c}$ baryons}}
	{\begin{tabular}{rcccclll}
			\toprule
			States	& Decay  & Partial & Total  & Exp.\cite{Workman:2022ynf} & Ref.\cite{PhysRevD.107.034031} & Ref.\cite{PhysRevD.96.116016}& Ref.\cite{PhysRevD.95.094008}\\
			&modes&width &width&width& & & \\
			&& (MeV)&(MeV)&(MeV)&(MeV) &(MeV) &(MeV) \\

			\hline
			
			$\Sigma_{c}^{++}(\frac{1}{2}^{+})$& $\Lambda_{c}\pi$   &2.12$\pm$0.16 &2.12$\pm$0.16 & 1.89$^{+0.09}_{-0.18}$ & 2$^{+1}_{-1}$& & \\
			$\Sigma_{c}^{+}(\frac{1}{2}^{+})$&  $\Lambda_{c}\pi$   & 2.61$\pm$0.20& 2.61$\pm$0.20&  2.30$\pm$0.40 & & & \\
			$\Sigma_{c}^{0}(\frac{1}{2}^{+})$&  $\Lambda_{c}\pi$   & 2.12$\pm$0.16& 2.12$\pm$0.16& 1.83$^{+0.11}_{-0.19}$ & & & \\
			$\Sigma_{c}^{*++}(\frac{3}{2}^{+})$& $\Lambda_{c}\pi$  & 15.95$\pm$1.24& 15.95$\pm$1.24& 14.78$^{+0.30}_{-0.40}$ & 15$^{+8}_{-8}$& & \\
			$\Sigma_{c}^{*+}(\frac{3}{2}^{+})$&  $\Lambda_{c}\pi$  &16.85$\pm$1.31 &16.85$\pm$1.31 & 17.20$^{+4.00}_{-2.20}$ & & & \\
			$\Sigma_{c}^{*0}(\frac{3}{2}^{+})$& $\Lambda_{c}\pi$  &15.95$\pm$1.24 &15.95$\pm$1.24 & 15.30$^{+0.40}_{-0.50}$ & & & \\
			
			$\Sigma_{c0}^{++}(\frac{1}{2}^{-})$&  $\Lambda_{c}\pi$   & 282.47$\pm$147.38& 282.47$\pm$147.38&  & 21$^{+9}_{-8}$& 22.65& 200\\	
			$\Sigma_{c0}^{+}(\frac{1}{2}^{-})$&  $\Lambda_{c}\pi$   &283.17$\pm$147.74 &283.17$\pm$147.74 &  & & & \\	
			$\Sigma_{c0}^{0}(\frac{1}{2}^{-})$&  $\Lambda_{c}\pi$   &282.47$\pm$147.38 &282.47$\pm$147.38 &   & & & \\	
			$\Sigma_{c1}^{++}(\frac{1}{2}^{-})$& $\Sigma_{c}\pi$   &123.24$\pm$45.47 &123.24$\pm$45.47 &  & 26$^{+12}_{-13}$& 17.63& 7.9\\
			$\Sigma_{c1}^{+}(\frac{1}{2}^{-})$& $\Sigma_{c}\pi$   &184.63$\pm$55.61 &184.63$\pm$55.61 &  & & & \\
			$\Sigma_{c1}^{0}(\frac{1}{2}^{-})$& $\Sigma_{c}\pi$  &123.24$\pm$45.47 &123.24$\pm$45.47 &   & & & \\		
			$\Sigma_{c1}^{++}(\frac{3}{2}^{-})$& $\Sigma_{c}^{*}\pi$   &1.88$\pm$0.31 &1.88$\pm$0.31 & 75$^{+22}_{-17}$ & 86$^{+40}_{-37}$& 36.5& \\
			$\Sigma_{c1}^{+}(\frac{3}{2}^{-})$& $\Sigma_{c}^{*}\pi$   & 2.79$\pm$0.37 & 2.79$\pm$0.37 & 62$^{+60}_{-40}$& & & \\
			$\Sigma_{c1}^{0}(\frac{3}{2}^{-})$& $\Sigma_{c}^{*}\pi$  & 1.88$\pm$0.31& 1.88$\pm$0.31& 72$^{+22}_{-15}$& & & \\	
			
			$\Sigma_{c2}^{++}(\frac{3}{2}^{-})$& $\Lambda_{c}\pi$   &99.51$\pm$23.14 &142.53$\pm$23.76 &   & 60$^{+25}_{-19}$& 24.69& 0.95\\
			&$\Sigma_{c}\pi$ &30.35$\pm$4.99 & & & & & \\
			&$\Sigma_{c}^{*}\pi$ &12.67$\pm$2.08 & & & & & \\
			$\Sigma_{c2}^{+}(\frac{3}{2}^{-})$& $\Lambda_{c}\pi$   &100.49$\pm$23.37 &164.77$\pm$24.28 &  & & & \\
			&$\Sigma_{c}\pi$ &45.37$\pm$6.09 & & & & & \\
			&$\Sigma_{c}^{*}\pi$ &18.91$\pm$2.54 & & & & & \\
			$\Sigma_{c2}^{0}(\frac{3}{2}^{-})$& $\Lambda_{c}\pi$   &99.51$\pm$23.14 &142.53$\pm$23.76 &  & & & \\
			&$\Sigma_{c}\pi$ &30.35$\pm$4.99 & & & & & \\
			&$\Sigma_{c}^{*}\pi$ &12.67$\pm$2.08 & & & & & \\
			
			$\Sigma_{c2}^{++}(\frac{5}{2}^{-})$& $\Lambda_{c}\pi$   &131.42$\pm$30.56 &185.33$\pm$31.23 &  & 164$^{+95}_{-86}$& 33.22& \\
			&$\Sigma_{c}\pi$ &20.57$\pm$3.38& & & & & \\
			&$\Sigma_{c}^{*}\pi$ &33.34$\pm$5.48 & & & & & \\
			
			$\Sigma_{c2}^{+}(\frac{5}{2}^{-})$&$\Lambda_{c}\pi$   &132.57$\pm$30.83 &213.15$\pm$31.82 &   & & & \\
			&$\Sigma_{c}\pi$ &30.77$\pm$4.13 & & & & & \\
			&$\Sigma_{c}^{*}\pi$ &49.81$\pm$6.69 & & & & & \\
			
			$\Sigma_{c2}^{0}(\frac{5}{2}^{-})$&$\Lambda_{c}\pi$   &131.42$\pm$30.56 &185.33$\pm$31.23 &  & & & \\
			&$\Sigma_{c}\pi$ &20.57$\pm$3.38 & & & & & \\
			&$\Sigma_{c}^{*}\pi$ &33.34$\pm$5.48 & & & & & \\
			\botrule          		& & & 
		\end{tabular}  \label{table3} }
\end{table}

The $1S$ $\Sigma_c$ baryons, such as  $\Sigma_{c}$ and $\Sigma_{c}^{*}$, are firmly established in experimental studies. Unlike ground state $\Lambda_{c}$ and $\Xi_{c}$, which decay weakly, $\Sigma_{c}$ and $\Sigma_{c}^{*}$ undergo strong decay via the $\Lambda_{c}\pi$ channel. Their decay widths are \cite{PhysRevD.75.014006}
 
\begin{equation}
	\label{eq:5}
	\begin{split}
		\Gamma[\Sigma_{c}^{++}(\frac{1}{2}^{+})]&=\Gamma[\Sigma_{c}^{++}(\frac{1}{2}^{+})\rightarrow\Lambda_{c}^{+}\pi^{+}]\\
		&=\frac{g_{2}^{2}}{2\pi f_{\pi}^{2}} \frac{M_{\Lambda_{c}^{+}}}{M_{\Sigma_{c}^{++}(\frac{1}{2}^{+})}}p_{\pi^{+}}^{3},
	\end{split}
\end{equation}

\begin{equation}
	\label{eq:6}
	\begin{split}
		\Gamma[\Sigma_{c}^{*++}(\frac{3}{2}^{+})]&=\Gamma[\Sigma_{c}^{*++}(\frac{3}{2}^{+})\rightarrow\Lambda_{c}^{+}\pi^{+}]\\
		&=\frac{g_{2}^{2}}{2\pi f_{\pi}^{2}} \frac{M_{\Lambda_{c}^{+}}}{M_{\Sigma_{c}^{*++}(\frac{3}{2}^{+})}}p_{\pi^{+}}^{3},
	\end{split}
 \end{equation}
where the coupling constant $g_{2}=0.591 \pm 0.023 $. Similar relations can be used to calculate the strong decay widths of the isospin partners of baryon listed in Eq. (\ref{eq:5}) and (\ref{eq:6}). Employing the theoretical masses $M_{th}$ derived from the RFT model,  we compute the strong decay width and present these results alongside the experimentally observed width  in Table \ref*{table3}. We observe that the experimental width of  $\Sigma_{c}$ and $\Sigma_{c}^{*}$ are well reproduced in HHChPT.

The strong decay width for $1P$-wave $\Sigma_{c}$ baryons in HHChPT are \cite{PhysRevD.75.014006}
\begin{equation}
	\label{eq:7}
	\begin{split}
		\Gamma[\Sigma_{c0}^{++}(\frac{1}{2}^{-})]&=\Gamma[\Sigma_{c0}^{++}(\frac{1}{2}^{-})\rightarrow\Lambda_{c}^{+}\pi^{+}]\\
		&=\frac{h_{3}^{2}}{2\pi f_{\pi}^{2}} \frac{M_{\Lambda_{c}^{+}}}{M_{\Sigma_{c0}^{++}(\frac{1}{2}^{-})}}E_{\pi^{+}}^{2}p_{\pi^{+}},
	\end{split}
\end{equation}

\begin{equation}
	\label{eq:8}
	\begin{split}
		\Gamma[\Sigma_{c1}^{++}(\frac{1}{2}^{-})]&=\Gamma[\Sigma_{c1}^{++}(\frac{1}{2}^{-})\rightarrow\Sigma_{c}^{++}\pi^{0}, \Sigma_{c}^{+}\pi^{+}]\\
		&=\frac{h_{4}^{2}}{4\pi f_{\pi}^{2}}\left(\frac{M_{\Sigma_{c}^{++}}}{M_{\Sigma_{c1}^{++}(\frac{1}{2}^{-})}}E_{\pi^{0}}^{2}p_{\pi^{0}}+\frac{M_{\Sigma_{c}^{+}}}{M_{\Sigma_{c1}^{++}(\frac{1}{2}^{-})}}E_{\pi^{+}}^{2}p_{\pi^{+}}	\right),
	\end{split}
\end{equation}

\begin{equation}
	\label{eq:9}
	\begin{split}
		\Gamma[\Sigma_{c1}^{++}(\frac{3}{2}^{-})]&=\Gamma[\Sigma_{c1}^{++}(\frac{3}{2}^{-})\rightarrow\Sigma_{c}^{*++}\pi^{0}, \Sigma_{c}^{*+}\pi^{+}]\\
		&=\frac{h_{9}^{2}}{9\pi f_{\pi}^{2}}\left(\frac{M_{\Sigma_{c}^{*++}}}{M_{\Sigma_{c1}^{++}(\frac{3}{2}^{-})}}p_{\pi^{0}}^{5}+\frac{M_{\Sigma_{c}^{*+}}}{M_{\Sigma_{c1}^{++}(\frac{3}{2}^{-})}}p_{\pi^{+}}^{5}	\right),
	\end{split}
\end{equation}

\begin{equation}
	\label{eq:10}
	\begin{split}
		\Gamma[\Sigma_{c2}^{++}(\frac{3}{2}^{-})]&=\Gamma[\Sigma_{c2}^{++}(\frac{3}{2}^{-})\rightarrow\Lambda_{c}^{+}\pi^{+},\Sigma_{c}^{++}\pi^{0}, \Sigma_{c}^{+}\pi^{+},\Sigma_{c}^{*++}\pi^{0}, \Sigma_{c}^{*+}\pi^{+}]\\
		&=\frac{4h_{10}^{2}}{15\pi f_{\pi}^{2}}\left(\frac{M_{\Lambda_{c}^{+}}}{M_{\Sigma_{c1}^{++}(\frac{3}{2}^{-})}}p_{\pi^{0}}^{5}\right)+\frac{h_{11}^{2}}{10\pi f_{\pi}^{2}}\left(\frac{M_{\Sigma_{c}^{++}}}{M_{\Sigma_{c2}^{++}(\frac{3}{2}^{-})}}p_{\pi^{0}}^{5}+\frac{M_{\Sigma_{c}^{+}}}{M_{\Sigma_{c2}^{++}(\frac{3}{2}^{-})}}p_{\pi^{+}}^{5}\right)\\
		&+\frac{h_{11}^{2}}{10\pi f_{\pi}^{2}}\left(\frac{M_{\Sigma_{c}^{*++}}}{M_{\Sigma_{c2}^{++}(\frac{3}{2}^{-})}}p_{\pi^{0}}^{5}+\frac{M_{\Sigma_{c}^{*+}}}{M_{\Sigma_{c2}^{++}(\frac{3}{2}^{-})}}p_{\pi^{+}}^{5} \right),
	\end{split}
\end{equation}

\begin{equation}
	\label{eq:11}
	\begin{split}
		\Gamma[\Sigma_{c2}^{++}(\frac{5}{2}^{-})]&=\Gamma[\Sigma_{c2}^{++}(\frac{5}{2}^{-})\rightarrow\Lambda_{c}^{+}\pi^{+},\Sigma_{c}^{++}\pi^{0}, \Sigma_{c}^{+}\pi^{+},\Sigma_{c}^{*++}\pi^{0}, \Sigma_{c}^{*+}\pi^{+}]\\
		&=\frac{4h_{10}^{2}}{15\pi f_{\pi}^{2}}\left(\frac{M_{\Lambda_{c}^{+}}}{M_{\Sigma_{c1}^{++}(\frac{5}{2}^{-})}}p_{\pi^{0}}^{5}\right)+\frac{2h_{11}^{2}}{45\pi f_{\pi}^{2}}\left(\frac{M_{\Sigma_{c}^{++}}}{M_{\Sigma_{c2}^{++}(\frac{5}{2}^{-})}}p_{\pi^{0}}^{5}+\frac{M_{\Sigma_{c}^{+}}}{M_{\Sigma_{c2}^{++}(\frac{5}{2}^{-})}}p_{\pi^{+}}^{5}\right)\\
		&+\frac{7h_{11}^{2}}{45\pi f_{\pi}^{2}}\left(\frac{M_{\Sigma_{c}^{*++}}}{M_{\Sigma_{c2}^{++}(\frac{5}{2}^{-})}}p_{\pi^{0}}^{5}+\frac{M_{\Sigma_{c}^{*+}}}{M_{\Sigma_{c2}^{++}(\frac{5}{2}^{-})}}p_{\pi^{+}}^{5}\right),
	\end{split}
\end{equation}
As per the quark model, coupling constants $h_{3}$,  $h_{4}$,  $h_{9}$,  $h_{10}$, and  $h_{11}$ are associated with $h_{2}$ or $h_{8}$ by \cite{PhysRevD.56.5483}
\begin{equation}
		\label{eq:11.1}
	\begin{split}
		&|h_{3}|=\sqrt{3}|h_{2}|, \hspace{1.189cm} |h_{4}|=2|h_{2}|, \\
		&|h_{8}|=|h_{9}|=|h_{10}|, \hspace{0.5cm}|h_{11}|=\sqrt{2}|h_{10}|.
	\end{split}
\end{equation}
Utilizing Eq. (\ref{eq:7})-(\ref{eq:11.1}) and the theoretical masses extracted from RFT model, the decay widths are calculated for $1P$-wave states of $\Sigma_{c}$ baryon and results are listed in Table \ref{table3}.

Currently, only one excited state, identified as $\Sigma_{c}(2800)$, has been found by the Belle\cite{Belle:2004zjl} and BABAR\cite{BaBar:2008get} Collaborations decaying through the $\Lambda_{c}\pi$ channel. Its spin-parity has not yet been determined. The mass calculations for $\Sigma_{c}$ baryons in RFT model predicts that $\Sigma_{c}(2800)$ corresponds to the $\Sigma_{c1}(\frac{3}{2}^{-})$ state. 
The calculated widths for the $1P$-wave states of the $\Sigma_{c}$ baryon indicate that among these five states, the $\Sigma_{c1}(\frac{3}{2}^{-})$ state is the narrowest. Additionally, the width of $\Sigma_{c}(2800)$ is the closest to the width of the $\Sigma_{c1}(\frac{3}{2}^{-})$ state among all five $1P$-wave states. Therefore, the assignment of $J^{P}=\frac{3}{2}^{-}$  for  $\Sigma_{c}(2800)$ is further supported by the strong decay analysis. Moreover, the broad width (several hundred MeV) for $1P$-wave states explains why they have not been experimentally observed yet.

\subsection{$\Xi_{c}^{'}$}
\begin{table}
	\tbl{{Same as in Table \ref{table2}, but for  $\Xi_{c}^{'}$ baryons.}}
	{\begin{tabular}{rcccccc}
			\toprule
			States	& Decay  & Partial & Total  & Exp.\cite{Workman:2022ynf} & Ref.\cite{PhysRevD.107.034031}& Ref.\cite{PhysRevD.96.116016}\\
			&modes&width &width&width& & \\
			&& (MeV)&(MeV)&(MeV)&(MeV) &(MeV) \\
			
			\hline
			
			$\Xi_{c}^{'+}(\frac{1}{2}^{+})$&Radiative &- &- & - & & \\
			$\Xi_{c}^{'0}(\frac{1}{2}^{+})$&Radiative &- &- & - & & \\		 
			$\Xi_{c}^{*+}(\frac{3}{2}^{+})$& $\Xi_{c}\pi$ &2.60$\pm$0.15 &2.60 & 2.14$\pm$0.19 & 0.4$^{+0.2}_{-0.2}$& \\	 
			$\Xi_{c}^{*0}(\frac{3}{2}^{+})$&$\Xi_{c}\pi$ &2.60$\pm$0.15 &2.60 & 2.35$\pm$0.22 & & \\
			$\Xi_{c0}^{'+}(\frac{1}{2}^{-})$& $\Lambda_{c}K$ &376.01$\pm$196.17 &828.57$\pm$257.60 && 7$^{+4}_{-3}$&  21.67\\
			&$\Xi_{c}\pi$ &452.56$\pm$166.96 & & & & \\
			$\Xi_{c0}^{'0}(\frac{1}{2}^{-})$& $\Lambda_{c}K$ &382.63$\pm$199.63 &835.19$\pm$260.24 && & \\
			&$\Xi_{c}\pi$ &452.56$\pm$ 166.96& & & & \\
			$\Xi_{c1}^{'+}(\frac{1}{2}^{-})$& $\Sigma_{c}K$ &0 &140.95$\pm$52.00 && 5$^{+2}_{-3}$& 37.05\\
			&$\Xi_{c}^{'}\pi$ &140.95$\pm$52.00 & & & & \\
			$\Xi_{c1}^{'0}(\frac{1}{2}^{-})$& $\Sigma_{c}K$&0 &140.95$\pm$ 52.00&7.10$\pm$2.00& & \\
			&$\Xi_{c}^{'}\pi$ & 140.95$\pm$52.00 & & & & \\
			$\Xi_{c1}^{'+}(\frac{3}{2}^{-})$& $\Sigma_{c}K$&0 &7.73$\pm$1.27 &15.00$\pm$9.00& 28$^{+14}_{-14}$&20.89 \\
			&$\Xi_{c}^{'}\pi$ & 7.73$\pm$1.27 & & & & \\
			$\Xi_{c1}^{'0}(\frac{3}{2}^{-})$& $\Sigma_{c}K$&0 &7.73$\pm$1.27 &10.20$\pm$1.40& & \\
			&$\Xi_{c}^{'}\pi$ &7.73$\pm$1.27 & & & & \\
			$\Xi_{c2}^{'+}(\frac{3}{2}^{-})$& $\Lambda_{c}K$&45.52$\pm$10.58 &199.41$\pm$23.27 & & 19$^{+9}_{-9}$&12.33 \\
			&$\Xi_{c}\pi$ &122.49$\pm$20.14 & & & & \\
			&$\Sigma_{c}K$ &1.79$\pm$0.24 & & & & \\ 
			&$\Xi_{c}^{'}\pi$ & 29.61$\pm$4.86& & & & \\			
			
			$\Xi_{c2}^{'0}(\frac{3}{2}^{-})$& $\Lambda_{c}K$&47.32$\pm$11.00 &200.57$\pm$23.46 & &  \\
			&$\Xi_{c}\pi$ &122.49$\pm$20.14 & & & & \\
			&$\Sigma_{c}K$ &1.15$\pm$0.19 & & & & \\
			&$\Xi_{c}^{'}\pi$ &29.61$\pm$4.86 & & & & \\

			$\Xi_{c2}^{'+}(\frac{5}{2}^{-})$& $\Lambda_{c}K$& 71.51$\pm$16.63 &266.41$\pm$32.78 && 43$^{+21}_{-21}$& 20.2 \\ 
			&$\Xi_{c}\pi$ & 170.52$\pm$28.04 & & & & \\
			&$\Sigma_{c}K$ & 3.91$\pm$0.53 & & & & \\
			&$\Xi_{c}^{'}\pi$ & 20.47$\pm$3.37 & & & & \\

			$\Xi_{c2}^{'0}(\frac{5}{2}^{-})$& $\Lambda_{c}K$&73.84$\pm$17.17 &267.39$\pm$33.05 && & \\
			&$\Xi_{c}\pi$ &170.52$\pm$28.04 & & & & \\
			&$\Sigma_{c}K$ &2.56$\pm$0.42 & & & & \\
			&$\Xi_{c}^{'}\pi$ &20.47$\pm$3.37 & & & & \\			
			
			\botrule          	
		\end{tabular} \label{table4}   }
\end{table}

The $\Xi_{c}^{'}$ baryonic family consists of established states in the $1S$-wave named $\Xi_{c}^{'}$ and $\Xi_{c}^{*}$. They were first reported by CLEO in $\Xi_{c}\gamma$\cite{CLEO:1998wvk} and $\Xi_{c}\pi$\cite{CLEO:1995amh}, respectively. Due to the small mass difference between $\Xi_{c}^{'}$ and $\Xi_{c}$ baryons, the strong decay $\Xi_{c}^{'}\rightarrow\Xi_{c}\pi$ is kinematically suppressed, restricting the decay modes to only $\Xi_{c}^{'}\rightarrow\Xi_{c}\gamma$. Meanwhile, $\Xi_{c}^{*}$ decays via strong interaction to $\Xi_{c}\pi$. It's decay width in HHChPT is \cite{PhysRevD.75.014006}
\begin{equation}
	\label{eq:12}
	\begin{split}
		\Gamma[\Xi_{c}^{*+}(\frac{3}{2}^{+})]&=\Gamma[\Xi_{c}^{*+}(\frac{3}{2}^{+})\rightarrow\Xi_{c}^{+}\pi^{0},\Xi_{c}^{0}\pi^{+}]\\
		&=\frac{g_{2}^{2}}{2\pi f_{\pi}^{2}} \left(\frac{1}{4}\frac{M_{\Xi_{c}^{+}}}{M_{\Xi_{c}^{*+}(\frac{3}{2}^{+})}}p_{\pi^{0}}^{3}+\frac{1}{2}\frac{M_{\Xi_{c}^{0}}}{M_{\Xi_{c}^{*+}(\frac{3}{2}^{+})}}p_{\pi^{+}}^{3}\right).
	\end{split}
\end{equation}
This gives $\Gamma[\Xi_{c}^{*+}(\frac{3}{2}^{+})]=2.6$ MeV, which is very close to the measured width of 2.04 MeV for $\Xi_{c}^{+}(2645)$. 

Further, the strong decay width of $1P$-wave states of $\Xi_{c}^{'+}$ in HHChPT are \cite{PhysRevD.95.094018}
\begin{equation}
	\label{eq:13}
	\begin{split}
		\Gamma[\Xi_{c0}^{'+}(\frac{1}{2}^{-})]&=\Gamma[\Xi_{c0}^{'+}(\frac{1}{2}^{-})\rightarrow\Lambda_{c}^{+}K^{0},\Xi_{c}^{+}\pi^{0},\Xi_{c}^{0}\pi^{+}]\\
		&=\frac{h_{3}^{2}}{2\pi f_{\pi}^{2}} \left(\frac{M_{\Lambda_{c}^{+}}}{M_{\Xi_{c0}^{'+}(\frac{1}{2}^{-})}}E_{K^{0}}^{2}p_{K^{0}}+\frac{1}{4}\frac{M_{\Xi_{c}^{+}}}{M_{\Xi_{c0}^{'+}(\frac{1}{2}^{-})}}E_{\pi^{0}}^{2}p_{\pi^{0}}+\frac{1}{2}\frac{M_{\Xi_{c}^{0}}}{M_{\Xi_{c0}^{'+}(\frac{1}{2}^{-})}}E_{\pi^{+}}^{2}p_{\pi^{+}}\right),
 	\end{split}
\end{equation}

\begin{equation}
	\label{eq:14}
	\begin{split}
		\Gamma[\Xi_{c1}^{'+}(\frac{1}{2}^{-})]&=\Gamma[\Xi_{c1}^{'+}(\frac{1}{2}^{-})\rightarrow\Xi_{c}^{'+}\pi^{0},\Xi_{c}^{'0}\pi^{+}]\\
		&=\frac{h_{4}^{2}}{4\pi f_{\pi}^{2}} \left(\frac{1}{4}\frac{M_{\Xi_{c}^{'+}}}{M_{\Xi_{c1}^{'+}(\frac{1}{2}^{-})}}E_{\pi^{0}}^{2}p_{\pi^{0}}+\frac{1}{2}\frac{M_{\Xi_{c}^{'0}}}{M_{\Xi_{c1}^{'+}(\frac{1}{2}^{-})}}E_{\pi^{+}}^{2}p_{\pi^{+}}\right),
	\end{split}
\end{equation}

\begin{equation}
	\label{eq:15}
	\begin{split}
		\Gamma[\Xi_{c1}^{'+}(\frac{3}{2}^{-})]&=\Gamma[\Xi_{c1}^{'+}(\frac{3}{2}^{-})\rightarrow\Xi_{c}^{'+}\pi^{0},\Xi_{c}^{'0}\pi^{+}]\\
		&=\frac{h_{9}^{2}}{9\pi f_{\pi}^{2}} \left(\frac{1}{4}\frac{M_{\Xi_{c}^{'+}}}{M_{\Xi_{c1}^{'+}(\frac{3}{2}^{-})}}p_{\pi^{0}}^{5}+\frac{1}{2}\frac{M_{\Xi_{c}^{'0}}}{M_{\Xi_{c1}^{'+}(\frac{3}{2}^{-})}}p_{\pi^{+}}^{5}\right),
	\end{split}
\end{equation}

\begin{equation}
	\label{eq:16}
	\begin{split}
		\Gamma[\Xi_{c2}^{'+}(\frac{3}{2}^{-})]&=\Gamma[\Xi_{c2}^{'+}(\frac{3}{2}^{-})\rightarrow\Lambda_{c}^{+}K^{0},\Xi_{c}^{+}\pi^{0},\Xi_{c}^{0}\pi^{+},\Sigma_{c}^{++}K^{-},\Sigma_{c}^{+}K^{0},\Sigma_{c}^{0}K^{+},\Xi_{c}^{'+}\pi^{0},\Xi_{c}^{'0}\pi^{+}]\\
		&=\frac{4h_{10}^{2}}{15\pi f_{\pi}^{2}}\left(\frac{M_{\Lambda_{c}^{+}}}{M_{\Xi_{c2}^{'+}(\frac{3}{2}^{-})}}p_{K^{0}}+\frac{M_{\Xi_{c}^{+}}}{M_{\Xi_{c2}^{'+}(\frac{3}{2}^{-})}}p_{\pi^{0}}+\frac{M_{\Xi_{c}^{0}}}{M_{\Xi_{c2}^{'+}(\frac{3}{2}^{-})}}p_{\pi^{+}}\right)\\
		&+\frac{h_{11}^{2}}{10\pi f_{\pi}^{2}}\left(\frac{M_{\Sigma_{c}^{++ }}}{M_{\Xi_{c2}^{'+}(\frac{3}{2}^{-})}}p_{K^{-}}^{5}+\frac{M_{\Sigma_{c}^{+ }}}{M_{\Xi_{c2}^{'+}(\frac{3}{2}^{-})}}p_{K^{0}}^{5}+\frac{M_{\Sigma_{c}^{0}}}{M_{\Xi_{c2}^{'+}(\frac{3}{2}^{-})}}p_{K^{+}}^{5}\right)\\
		&+\frac{h_{11}^{2}}{10\pi f_{\pi}^{2}} \left(\frac{M_{\Xi_{c}^{'+}}}{M_{\Xi_{c2}^{'+}(\frac{3}{2}^{-})}}p_{\pi^{0}}^{5}+\frac{M_{\Xi_{c}^{'0}}}{M_{\Xi_{c2}^{'+}(\frac{3}{2}^{-})}}p_{\pi^{+}}^{5}\right),
	\end{split}
\end{equation}

\begin{equation}
	\label{eq:17}
	\begin{split}
		\Gamma[\Xi_{c2}^{'+}(\frac{5}{2}^{-})]&=\Gamma[\Xi_{c2}^{'+}(\frac{5}{2}^{-})\rightarrow\Lambda_{c}^{+}K^{0},\Xi_{c}^{+}\pi^{0},\Xi_{c}^{0}\pi^{+},\Sigma_{c}^{++}K^{-},\Sigma_{c}^{+}K^{0},\Sigma_{c}^{0}K^{+},\Xi_{c}^{'+}\pi^{0},\Xi_{c}^{'0}\pi^{+}]\\
		&=\frac{4h_{10}^{2}}{15\pi f_{\pi}^{2}}\left(\frac{M_{\Lambda_{c}^{+}}}{M_{\Xi_{c2}^{'+}(\frac{5}{2}^{-})}}p_{K^{0}}+\frac{M_{\Xi_{c}^{+}}}{M_{\Xi_{c2}^{'+}(\frac{5}{2}^{-})}}p_{\pi^{0}}+\frac{M_{\Xi_{c}^{0}}}{M_{\Xi_{c2}^{'+}(\frac{5}{2}^{-})}}p_{\pi^{+}}\right)\\
		&+\frac{2h_{11}^{2}}{45\pi f_{\pi}^{2}}\left(\frac{M_{\Sigma_{c}^{++ }}}{M_{\Xi_{c2}^{'+}(\frac{5}{2}^{-})}}p_{K^{-}}^{5}+\frac{M_{\Sigma_{c}^{+ }}}{M_{\Xi_{c2}^{'+}(\frac{5}{2}^{-})}}p_{K^{0}}^{5}+\frac{M_{\Sigma_{c}^{0}}}{M_{\Xi_{c2}^{'+}(\frac{5}{2}^{-})}}p_{K^{+}}^{5}\right)\\
		&+\frac{2h_{11}^{2}}{45\pi f_{\pi}^{2}} \left(\frac{M_{\Xi_{c}^{'+}}}{M_{\Xi_{c2}^{'+}(\frac{5}{2}^{-})}}p_{\pi^{0}}^{5}+\frac{M_{\Xi_{c}^{'0}}}{M_{\Xi_{c2}^{'+}(\frac{5}{2}^{-})}}p_{\pi^{+}}^{5}\right),
	\end{split}
\end{equation}
where $p_{K}$ and $E_{K}$ denote the kaon's momentum and energy in the center of mass frame, respectively. Same approach can be used to calculate width for the $1P$ states of $\Xi_{c}^{'0}$ baryon as well. The results obtained from these calculations are presented in Table \ref{table4}. Observations from the BABAR \cite{PhysRevD.77.031101} and Belle \cite{Belle:2017jrt,Belle:2018yob} Collaborations  identified the $\Xi_{c}(2930)$ state in   $\Xi_{c}\pi$ channel. Subsequently, LHCb Collaboration\cite{LHCb:2020iby} announced observation of two states, $\Xi_{c}(2923)$ and $\Xi_{c}(2939)$, in same $\Xi_{c}\pi$ channel.  It is possible that the peak observed for $\Xi_{c}(2930)$ by BABAR and Belle could be resolved into two distinct peaks, corresponding to $\Xi_{c}(2923)$ and $\Xi_{c}(2939)$, by LHCb. The mass of $\Xi_{c1}^{'}(\frac{3}{2}^{-})$ state, as predicted by the RFT model, along with the strong decay width determined in HHChPT, aligns excellently  with the measured masses and width of  $\Xi_{c}(2930)$ by Belle or $\Xi_{c}(2939)$ by LHCb. Hence, the $J^{P}=\frac{3}{2}^{-}$ assignment for  $\Xi_{c}(2930)/\Xi_{c}(2939)$ is corroborated by both mass spectra and strong decay analysis.
Furthermore, the calculated mass in RFT model for $\Xi_{c1}^{'}(\frac{1}{2}^{-})$ is close to mass  for $\Xi_{c}(2923)$ with difference of 36 MeV. However, the calculated width of 140.95 MeV for  $\Xi_{c1}^{'}(\frac{1}{2}^{-})$ is significantly larger than the measured width of 7.10 MeV for $\Xi_{c}(2923)$, thereby reducing the likelihood of $\Xi_{c1}^{'}(\frac{1}{2}^{-})$ being identified as $\Xi_{c}(2923)$.

\subsection{$\Omega_{c}$}
\begin{table}
	\tbl{{Same as in Table \ref{table2}, but for $\Omega_{c}$ baryons}}
	{\begin{tabular}{rccccll}
			\toprule
			States	& Decay  & Partial & Total  & Ref.\cite{PhysRevD.107.034031} & Ref.\cite{PhysRevD.95.116010}& Ref.\cite{PhysRevD.95.094008}\\
			&modes&width &width& & & \\
			&& (MeV)&(MeV)&(MeV) & (MeV)&(MeV) \\
			\hline
			$\Omega_{c}^{0}(\frac{1}{2}^{+})$&Weak &-&-& & & \\
			$\Omega_{c}^{*0}(\frac{3}{2}^{+})$&Radiative &-&-& & & \\
			$\Omega_{c0}^{0}(\frac{1}{2}^{-})$& $\Xi_{c}K$ &423.12$\pm$156.15&423.12$\pm$156.15&4$^{+2}_{-2}$ & & 820\\
			$\Omega_{c1}^{0}(\frac{1}{2}^{-})$& $\Xi_{c}^{'}K$&0&0&8$^{+4}_{-4}$ & &29 \\
			$\Omega_{c1}^{0}(\frac{3}{2}^{-})$& $\Xi_{c}^{'}K$&0&0&26$^{+13}_{-13}$ &  0.94 & \\
			$\Omega_{c2}^{0}(\frac{3}{2}^{-})$& $\Xi_{c}K$
			&33.05$\pm$5.43 &33.61$\pm$5.44 & 7$^{+3}_{-3}$& 4.96& \\
			&$\Xi_{c}^{'}K$ &0.56$\pm$0.09 & & & & \\
			$\Omega_{c2}^{0}(\frac{5}{2}^{-})$& $\Xi_{c}K$
			&63.47$\pm$10.44 &65.35$\pm$10.44 &50$^{+25}_{-24}$ &9.53 & \\
			&$\Xi_{c}^{'}K$ &1.88$\pm$0.31 & & & & \\
			\botrule          		& 
		\end{tabular}  \label{table5}  }
\end{table}

For the $\Omega_{c}$ baryonic family, the  $1S$ states named $\Omega_{c}^{0}$ and $\Omega_{c}^{0}(2770)$ are well established. The ground state,  $\Omega_{c}^{0}$, decays weakly. While the $\Omega_{c}^{0}(2770)$ was first observed in the radiative  decay channel $\Omega_{c}\gamma$  by BABAR \cite{BaBar:2006pve}. Furthermore, the LHCb observed five narrow excited states of $\Omega_{c}$ baryon in $\Xi_{c}^{+}\pi^{-}$ channel, named $\Omega_{c}^{0}(3000)$, $\Omega_{c}^{0}(3050)$, $\Omega_{c}^{0}(3066)$, $\Omega_{c}^{0}(3090)$, and $\Omega_{c}^{0}(3119)$, whose width were measured to be $4.5\pm0.7$ MeV, $<1.2$ MeV, $3.3 \pm0.6$ MeV, $8.7 \pm1.3$ MeV, and $<2.6$ MeV, respectively \cite{LHCb:2017uwr}. The spin-parity of these five states are unknown.

In the framework of HHChPT, the strong decay width expressions derived from the relevant chiral Lagrangian for the $1P$ states are \cite{PhysRevD.95.094018}
\begin{equation}
	\label{eq:18}
	\begin{split}
		\Gamma[\Omega_{c0}^{0}(\frac{1}{2}^{-})]&=\Gamma[\Omega_{c0}^{0}(\frac{1}{2}^{-})\rightarrow\Xi_{c}^{0}K^{0},\Xi_{c}^{+}K^{-}]\\
		&=\frac{h_{3}^{2}}{2\pi f_{\pi}^{2}} \left(\frac{M_{\Xi_{c}^{0}}}{M_{\Omega_{c0}^{0}(\frac{1}{2}^{-})}}E_{K^{0}}^{2}p_{K^{0}}+\frac{M_{\Xi_{c}^{+}}}{M_{\Omega_{c0}^{0}(\frac{1}{2}^{-})}}E_{K^{-}}^{2}p_{K^{-}}\right),
	\end{split}
\end{equation}

\begin{equation}
	\label{eq:19}
	\begin{split}
		\Gamma[\Omega_{c1}^{0}(\frac{1}{2}^{-})]&=\Gamma[\Omega_{c1}^{0}(\frac{1}{2}^{-})\rightarrow\Xi_{c}^{'0}K^{0},\Xi_{c}^{'+}K^{-}]\\
		&=\frac{h_{4}^{2}}{4\pi f_{\pi}^{2}} \left(\frac{M_{\Xi_{c}^{'0}}}{M_{\Omega_{c1}^{0}(\frac{1}{2}^{-})}}E_{K^{0}}^{2}p_{K^{0}}+\frac{M_{\Xi_{c}^{'+}}}{M_{\Omega_{c1}^{0}(\frac{1}{2}^{-})}}E_{K^{-}}^{2}p_{K^{-}}\right),
	\end{split}
\end{equation}

\begin{equation}
	\label{eq:20}
	\begin{split}
		\Gamma[\Omega_{c1}^{0}(\frac{3}{2}^{-})]&=\Gamma[\Omega_{c1}^{0}(\frac{3}{2}^{-})\rightarrow\Xi_{c}^{'0}K^{0},\Xi_{c}^{'+}K^{-}]\\
		&=\frac{h_{9}^{2}}{9\pi f_{\pi}^{2}} \left(\frac{M_{\Xi_{c}^{'0}}}{M_{\Omega_{c1}^{0}(\frac{3}{2}^{-})}}p_{K^{0}}^{5}+\frac{M_{\Xi_{c}^{'+}}}{M_{\Omega_{c1}^{0}(\frac{3}{2}^{-})}}p_{K^{-}}^{5}\right),
	\end{split}
\end{equation}

\begin{equation}
	\label{eq:21}
	\begin{split}
		\Gamma[\Omega_{c2}^{0}(\frac{3}{2}^{-})]&=\Gamma[\Omega_{c2}^{0}(\frac{3}{2}^{-})\rightarrow\Xi_{c}^{0}K^{0},\Xi_{c}^{+}K^{-},\Xi_{c}^{'0}K^{0},\Xi_{c}^{'+}K^{-}]\\
		&=\frac{4h_{10}^{2}}{15\pi f_{\pi}^{2}} \left(\frac{M_{\Xi_{c}^{0}}}{M_{\Omega_{c2}^{0}(\frac{3}{2}^{-})}}p_{K^{0}}^{5}+\frac{M_{\Xi_{c}^{+}}}{M_{\Omega_{c2}^{0}(\frac{3}{2}^{-})}}p_{K^{-}}^{5}\right)\\
		&+\frac{h_{11}^{2}}{10\pi f_{\pi}^{2}} \left(\frac{M_{\Xi_{c}^{'0}}}{M_{\Omega_{c2}^{0}(\frac{3}{2}^{-})}}p_{K^{0}}^{5}+\frac{M_{\Xi_{c}^{'+}}}{M_{\Omega_{c2}^{0}(\frac{3}{2}^{-})}}p_{K^{-}}^{5}\right),
	\end{split}
\end{equation}

\begin{equation}
	\label{eq:22}
	\begin{split}
		\Gamma[\Omega_{c2}^{0}(\frac{5}{2}^{-})]&=\Gamma[\Omega_{c2}^{0}(\frac{5}{2}^{-})\rightarrow\Xi_{c}^{0}K^{0},\Xi_{c}^{+}K^{-},\Xi_{c}^{'0}K^{0},\Xi_{c}^{'+}K^{-}]\\
		&=\frac{4h_{10}^{2}}{15\pi f_{\pi}^{2}} \left(\frac{M_{\Xi_{c}^{0}}}{M_{\Omega_{c2}^{0}(\frac{5}{2}^{-})}}p_{K^{0}}^{5}+\frac{M_{\Xi_{c}^{+}}}{M_{\Omega_{c2}^{0}(\frac{5}{2}^{-})}}p_{K^{-}}^{5}\right)\\
		&+ \frac{2h_{11}^{2}}{45\pi f_{\pi}^{2}} \left(\frac{M_{\Xi_{c}^{'0}}}{M_{\Omega_{c2}^{0}(\frac{5}{2}^{-})}}p_{K^{0}}^{5}+\frac{M_{\Xi_{c}^{'+}}}{M_{\Omega_{c2}^{0}(\frac{5}{2}^{-})}}p_{K^{-}}^{5}\right).
	\end{split}
\end{equation}
The calculated strong decay widths are presented in Table \ref{table5}. 
The mass spectra calculated in RFT approach predicts  $\Omega_{c}(3000)$, $\Omega_{c}(3050)$, $\Omega_{c}(3065)$, $\Omega_{c}(3090)$, and $\Omega_{c}(3119)$ baryons to belong to $\Omega_{c0}(\frac{1}{2}^{-})$, $\Omega_{c1}(\frac{1}{2}^{-})$, $\Omega_{c1}(\frac{3}{2}^{-})$, $\Omega_{c2}(\frac{3}{2}^{-})$, and $\Omega_{c2}(\frac{5}{2}^{-})$, respectively.  However, we observe a substantial strong decay width of 423.12 MeV for the 
 $\Omega_{c0}(\frac{1}{2}^{-})$ state. Additionally, for the $\Omega_{c1}(\frac{1}{2}^{-})$ and $\Omega_{c1}(\frac{3}{2}^{-})$ states, the  $\Xi_{c}^{'}K$ channel is the only allowed decay mode since the $\Xi_{c}K$ channel is restricted under the heavy quark limit. In our model, the $\Xi_{c}^{'}K$ channel is also suppressed due to phase space constraints. Furthermore, the calculated widths for the $\Omega_{c2}(\frac{3}{2}^{-})$ and $\Omega_{c2}(\frac{5}{2}^{-})$ states are considerably larger than the measured widths of the five excited $\Omega_{c}$ baryon resonances. Therefore, it is challenging to assign specific spin-parity values to these five observed $\Omega_{c}$ resonances based solely on strong decay calculations. This highlights the need for additional theoretical and experimental investigations to accurately identify these states.

\section{Conclusion}

The strong decays of $1S$ and $1P$ states of singly charmed baryons into another singly charmed baryon and a light pseudoscalar meson are analyzed using heavy hadron chiral perturbation theory (HHChPT). This approach accurately reproduces the strong decay widths of well-established singly charmed baryon states. Based on the mass spectra from the relativistic flux tube (RFT) model and strong decays in HHChPT, we have investigated the spin-parity assignments. The $\Sigma_{c}(2800)$ and $\Xi_{c}(2939)$ are identified to be the states belonging to the $1P$-wave with $J^{P}=\frac{3}{2}^{-}$. However, we cannot definitively determine the spin-parity assignments for  $\Omega_{c}(3000)$, $\Omega_{c}(3050)$, $\Omega_{c}(3065)$, $\Omega_{c}(3090)$, and $\Omega_{c}(3119)$ baryons  because the calculated widths for $|1P, 1/2^-\rangle_{j=0}$ is too broad to be observed, while the widths for $|1P, 3/2^-\rangle_{j=2}$ and  $|1P, 5/2^-\rangle_{j=2}$ are larger than the measured widths of these five resonances. Additionally, the strong decay channels for the remaining two states in the $1P$-wave are suppressed due to phase space constraints in our model. Therefore, further theoretical and experimental research is needed to identify these five $\Omega_{c}$ baryon resonances.
	
	\section*{Acknowledgments}
	The authors thank the organisers of the 12th International Conference on New Frontiers in Physics (ICNFP 2023) for the opportunity to present our work. Ms. Pooja Jakhad acknowledges the financial support provided by the Council of Scientific \& Industrial Research (CSIR) through the JRF-FELLOWSHIP scheme, file number 09/1007(13321)/2022-EMR-I.
	
	\bibliographystyle{ws-ijmpa}
	\bibliography{sample}
\end{document}